# Towards a Sociolinguistics-Based Framework for the Study of Politeness in Human-Computer Interaction

Ella Bar-Or, Tom Regev, Paz Shaviv, Noam Tractinsky[1]

Software and Information Systems Engineering
Ben-Gurion University of the Negev

**ABSTRACT**

Politeness plays an important role in regulating communication and enhancing social interactions. Research suggests that people treat interactive systems as social agents and may expect those systems to exhibit polite behavior. We augment early research in this area by proposing a framework that is grounded in sociolinguistics and pragmatics. The framework focuses on two complementary concepts – clarity and politeness. We suggest that these concepts are pertinent to various domains of interactive technologies and provide examples for the applicability of clarity and politeness rules to human-computer interaction. We conducted a laboratory experiment, in which politeness and clarity served as independent factors in the context of a cooperative computer game, based on the *Peekaboom* game. The manipulation of clarity failed, yet politeness was manipulated successfully based on the framework's rules of politeness. The results provide empirical support for the basic propositions of the framework and may facilitate more systematic research on politeness in various domains of social computing and other areas related to HCI.

**Keywords:** *Politeness in HCI, polite computing, Peekaboom game.*

## 1. INTRODUCTION

Politeness is considered a manifestation of human civilization, and one of the most effective strategies to modulate interpersonal relationships in human communication (Hau and Chi, 2013). As a commonsense term, it implies "proper behavioral conduct" (Eelen, 2014). Scholastic treatment of the concept mostly emphasizes the preservation of positive relations among people. By far, politeness research has been dominated by sociolinguistic theories. Thus, Brown and Levinson (1987) suggest that politeness seeks to establish a positive relationship between parties, and Lakoff (1990) defined politeness as "[…] a system of interpersonal relations designed to facilitate interaction by minimizing the potential for conflict and confrontation inherent in all human interchange" (p. 34).

Early research on politeness in human-computer interaction (HCI) can be traced to Reeves and Nass's (1996) seminal Media Equation theory, in which people were shown to treat computers in a manner that resembles politeness aspects of interpersonal behavior. Later HCI research has recognized the potential importance of politeness in the reverse direction, that is in terms of how computers treat users (e.g., Nass, 2004; Whitworth and Liu, 2010; Hayes and Miller, 2010). Such a perspective has implications for the design and use of interactive technology. Studies have suggested that polite behavior is so ingrained in human-to-human interaction that it transcends this domain into the realm of people's interactions with various types of media (e.g., Hayes et al., 2002; Nass, 2004). However, politeness in HCI still remains an underdeveloped area in terms of both theory and evidence.

---

[1] Corresponding author: noamt@bgu.ac.il

We propose a more systematic treatment of politeness in HCI by bringing to bear existing social science frameworks that suggest rules for efficient and resilient communication between people. We see this as a starting point for a research program with three main goals. First, we would like to explore the conceptualization of polite computing based on its treatment in the fields of pragmatics and sociolinguistics. Second, we are interested in establishing a research framework that could be generalized to different HCI domains to which the concept may apply, e.g., traditional HCI, computer-mediated communication (CMC), and human-robot interaction (HRI). Third, there is a need to empirically demonstrate the framework's viability in affecting users' perceptions and in supporting experimental manipulations of the politeness construct.

We begin by reviewing the literature on politeness in HCI (Section 2) and then move to sociolinguistics theories of politeness and cooperation and their relevance to HCI (Section 3). In Section 4, we propose a research framework. The following sections describe an experiment conducted to assess the framework and discuss the results and their implications.

## 2. POLITENESS IN HUMAN-COMPUTER INTERACTION

Communication between two parties has two major objectives: promote mutual understanding and improve relationships between the parties involved (Te'eni, 2001). These objectives can be fulfilled using various communication strategies, One such strategy involves controlling the communication process to improve its effectiveness. Other strategies relate to providing affective content and considering the receiver's perspective. Although Te'eni's work does not explicitly include politeness as a strategy, it offers a comprehensive framework for studying, exploring, and designing polite interactive technology. Thus, communication elements of control and relationship can be found implicitly in the two main streams of research on politeness in HCI, namely Media Equation Theory and Polite Computing. Whereas Media Equation Theory appears to emphasize relationship communication strategies, Polite Computing seems to address a mix of control and relationship communication strategies.

### 2.1 Media Equation Theory

One of the earliest considerations of politeness in the broad context of HCI is reported in Reeves and Nass's (1996) Media Equation Theory, which proposes that people respond to computer-based media as they do to other people. For example, people are unwilling and are hesitant to hurt the feelings of another person (or computer), so they tend to treat them more politely (Nass et al., 1999) Nass and Moon (2000) suggest that such behavior stems from overlearning social rules, as people fail to recognize that, when interacting with computers, the basis for applying the rules (e.g., not hurting people's feelings) no longer exists.

Nass (2004) proposes an evolutionary psychology-based explanation to this phenomenon. Once the politeness "script" is initiated, people stop searching for additional context cues and simply respond according to the script (Nass and Moon, 2000), whether the script is initiated by a conversation with another human or with a computer. Moreover, Reeves and Nass (1996) extrapolated, based on the norm of reciprocity (Gouldner, 1960), that we also expect computers to return the favor and be as polite to us as we are to them. Therefore, not only are we polite towards computers, but we also expect computers to be polite to us. "The biggest reason for making machines that are polite to people is that people are polite to machines. Everyone expects reciprocity, and everyone will be disappointed if it's absent" (Reeves and Nass, 1996, p.28-29). Nass and his colleagues have mostly demonstrated that humans behave politely to various forms of media, including computers. However, their work has been less explicit regarding how people would react to various aspects of politeness exhibited by interactive technology. Yet, they suggest (Reeves and Nass, 1996), as a general direction, that to support pleasant interactions designers should follow Grice's maxims on human communication (Grice, 1975). We will elaborate on this direction in Section 3.

## 2.2 Polite Computing

A subsequent stream of research can be traced to an edited volume titled "Human-Computer Etiquette" (Hayes and Miller, 2010), in which the terms etiquette and politeness had been used interchangeably. Hayes et al. (2002) defined HCI etiquette rules as a type of software design guideline aimed at facilitating smooth and effective interactions between humans and computers. In coining the term "Polite Computing," Whitworth (2005) distinguished etiquette (specific actions that may vary between cultures) from politeness (a general, common goal). According to Whitworth, the essence of politeness is offering others a choice and not taking it away from them. By reviewing impolite software and the deleterious effects of software rudeness, Whitworth has highlighted the need for specifying politeness guidelines for HCI design. He suggests that software politeness can be modeled based on Miller's core etiquette question: ''If this system were replaced by an ideal human assistant, … how would that assistant behave?'' (Miller, 2004, p. 33). Whitworth (2005) suggests four rules of person-to-person politeness that can serve as design principles for human-computer politeness.

*Respectfulness* - not taking another's rightful choices. Polite software should not preempt rightful user information choices regarding common HCI resources. For example, software should not change a browser home page without asking for permission. Similarly, program installation should ask before installing a new feature or application (Whitworth and Liu, 2008).

*Openness* - part of a polite greeting in most cultures is to introduce oneself and state one's business. Polite software should do the same. Users should see who is doing what on their computer. For example, users are more likely to trust a system if they know why it is doing what it is doing and this rarely happens when the software automatically runs its processes (Whitworth and Liu, 2008).

*Helpfulness* - polite software should help the user by offering understandable choices, as users cannot properly choose from options they do not understand. For example, explanations provided by the system must be at the appropriate level of abstraction regarding the user's domain knowledge. This is true whether it interacts with an expert user (in which case helpfulness means using words parsimoniously to facilitate efficient communication) or with a simple user (in which case helpfulness requires detailed explanations to facilitate clear understanding and learning) (Hayes et al., 2002).

*Be personal* - polite software should be aware of what the user already knows and remember users' choices and past interactions. It is inconsiderate to ignore past responses because that causes the user to redo them. Being personal also means that the software should adapt to the individual where possible (Whitworth, 2005).

## 2.3 Extending the Politeness Principles beyond Traditional Software

Besides interactions with traditional software, politeness can also benefit other HCI areas. Computer-mediated communication is an area naturally suited for politeness considerations. Obviously, users can benefit from using a platform that supports polite communication. For instance, in mobile phone calls, recipients often face a constant dilemma between ignoring calls at the possible expense of offending the caller on one hand and answering them at the risk of appearing rude and impolite towards others with whom they share a social activity, on the other hand (Inbar et al., 2014). Hence, technology-mediated social interaction should be designed to help users maintain politeness while handling the dilemma, e.g., by providing mechanisms that decrease distracting interruptions. Similarly, Whitworth and Liu (2010) proposed an etiquette-based design to address email spam.

HRI is another domain for which politeness has become increasingly relevant. There is a growing awareness of the need to study social aspects of HRI, with indications that in many cases, phenomena found in human social behavior also apply to situations in which computers and robots mesh into the social context. For example, Stein and Ohler (2017) suggest that the uncanny valley model, usually attributed to the physical appearance of human-like characters, can also be observed when it comes to the social behavior of such characters. Huisman (2017) surveyed the development of social touch technology and its potential effects on human recipients of the touch. Obviously, being touched by others has social ramifications, including on aspects of politeness. Indeed, findings suggest a relationship between an agent's (e.g., robot) level of politeness and the user's perception of the agent's politeness (Inbar and Meyer, 2019; Kato et al., 2015).

The basic notion of politeness, described by the Media Equation Theory, and Whitworth's four principles, provides a good starting point for research on politeness in HCI. However, given the increasing role of politeness in a large variety of computing-related areas, we need a more intricate and systematic research framework. In the next section, we propose such a framework based on politeness theories from the field of sociolinguistics.

## 3. SOCIOLINGUISTICS APPROACHES TO POLITENESS AND THEIR APPLICATION TO HCI

Our review of the literature suggests that while HCI politeness research has its roots in the social sciences, politeness mostly serves as a synonym for other ill-defined concepts such as etiquette and manners, and manipulations of politeness appear intuitive rather than theory-based. In this work, we attempt to borrow from social science theories of politeness, which offer rich and multifaceted treatment of the concept. Theorizing about politeness has mostly taken place within the academic fields of pragmatics and sociolinguistics. Most notable among those theories are Lakoff's (1973) and Leech's (1983) principles and Brown and Levinson's Model of politeness (1987). Importantly, all those approaches to politeness have emerged dialectically from Grice's (1967/1975) work on the cooperative principles and the rules of conversation.

Some of the chapters in Hayes and Miller's (2010) edited book (Johnson and Wang, 2010; Whitworth, 2005; Wu et al., 2010) have adopted Brown and Levinson's (1987) theory of politeness, which centers on the concept of "face", to guide research in this area. However, in formulating polite computing's rules, Whitworth (2005) does not expound on how the four rules emanate from that concept. In addition, we believe that despite its popularity, the concept of face and the politeness strategies that surround it may not be particularly suitable for HCI for several reasons. First, the concept of *face* may be too strong for the context of human-computer interaction. Second, face-saving strategies rely on verbal communication whereas we look for a framework that also addresses nonverbal actions. Third, the face concept might be susceptible to inter-cultural differences (Kasper, 1990; Mao, 1994; Eelen, 2014). Finally, Brown and Levinson's theory is relatively complex, and translating its rules to actionable HCI design rules might be less straightforward.

Following our review of these major works, we found the approaches of Grice and Lakoff to be the most promising for our goals. First, these theories cover distinctively different, yet complementary, points of view. Grice's Cooperative Principle mostly presents rational goals of communication and coordination, whereas Lakoff's politeness principles, which also aim at keeping the conversation afloat, stress, in addition, interpersonal relations and emotional considerations, which are not at the focus of Grice's treatment of the subject. In Te'eni's (2001) communication model, these points of view reflect the objectives of mutual understanding (mainly by controlling the process) and relationship (mainly by considering the other), respectively. Second, both of these theories have specific principles that can be applied to HCI. Finally, another requisite of politeness theory in HCI is that it covers not only linguistic aspects of communication but also behavioral ones. Grice's and Lakoff's approaches provide principles that can also be manifested in behavioral terms. Below, we present those principles and provide examples for how those principles can be adhered to (or ignored) in various HCI contexts.

### 3.1 Grice's Cooperative Principle

The roots of virtually all of social sciences' treatments of politeness can be found in Grice's seminal work (1967/1975) on the principles of human communication. Grice's theory rests on the assumption that conversationalists are rational individuals, who are, all other things being equal, primarily interested in the efficient conveyance of messages. To this end, Grice proposed the general Cooperative Principle (CP): "Make your conversational contribution such as is required, at the stage at which it occurs, by the accepted purpose or direction of talk exchange in which you are engaged" (Grice, 1975; p. 45). The CP subsumes four conversational categories (often also referred to as rules or maxims).

*The Maxim of Quantity*, where one tries to be as informative as one possibly can, providing as much information as needed, neither more nor less. For instance, responding to a simple yes/no question with a long monologue would violate the maxim. However, when a software program interprets a user action as an error, it should not only state the fact that an error occurred but also has to add informative suggestions on how the user can recover this time and avoid doing invalid action next time. Likewise, in a social network interaction, the software should not prevent the exchange of pertinent information when it comes to facilitating communication. For example, when responding to an online invitation, the system should support not only an Accept/Reject option but also the addition of a short explanation, in case one cannot accept the invitation.

*The Maxim of Quality* refers to one being truthful, and not giving information that is false or that is not supported by evidence. For example, if the system informs the user that downloading a movie will take about three minutes, but eventually it takes ten, the truthfulness of the system's original notification comes into question and may annoy the user. When it comes to CMC (e.g., social networks) or HRI, presenting the correct information is critical, as this aspect of communication is crucial in domains that emphasize relationships.

*The Maxim of Manner* refers to being as clear, as brief, and as orderly as one can in what one says, and to avoid obscurity and ambiguity. The system's interactions with a user should thus be as clear as possible. For instance, using overly technical language, i.e., not speaking the user's language (Molich and Nielsen, 1990) should be discouraged. The rule of manner can be generalized to interface design by requiring that the interface present only necessary controls and do so in an unambiguous manner. A disorganized interface, too many controls, links, or other call-to-actions cause confusion. Ambiguity in social networks might arise, for example, when a social application suggests a new friend (e.g., "people you may know") without presenting the reason for that suggestion. Consider a connection request on LinkedIn. Until recently, it was very difficult for an inviting person to indicate to the invitee how they may know each other. Consequently, invitations were done inappropriately and confused the invitees (Figure 1, top panel). The invitation was later redesigned to include a handy mechanism supporting the inviting person in adding a personal note that can clarify how he or she might be related to the invitee (Figure 1, bottom panel).

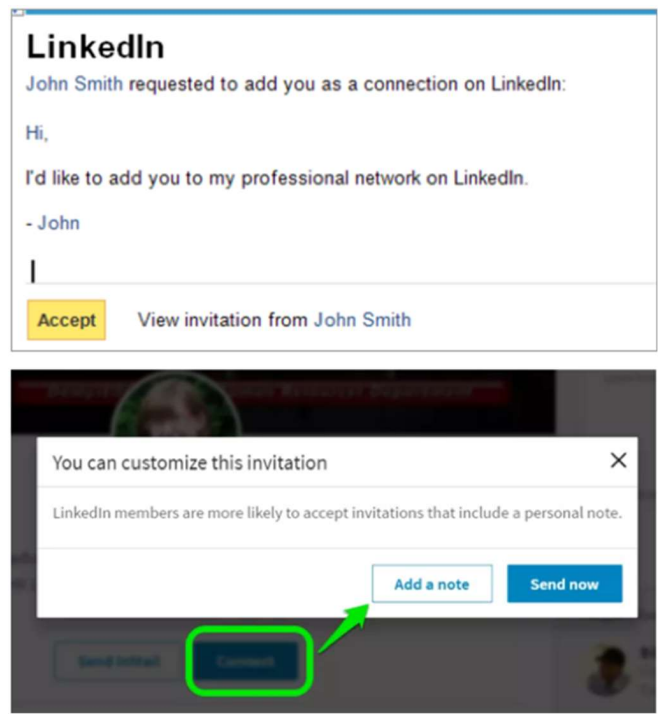

**Figure 1.** Top: the original invitation to connect with other LinkedIn members did not offer a customization option. Bottom: the redesigned invitation does prompt users to customize the invitation.

*The Maxim of Relation* calls for conversationalists to make a relevant contribution depending on the context and needs of the other person. This rule can be instantiated, for example, by adapting the system's messages to the user's level of domain expertise (e.g., Tractinsky et al., 1998). More generally, the rule suggests that designers should consider users' needs and save them unnecessary trouble and effort. One such example is an organizational email system that informs a sender before she even starts composing a new mail message that the addressee will be unavailable until a certain date (Figure 2).

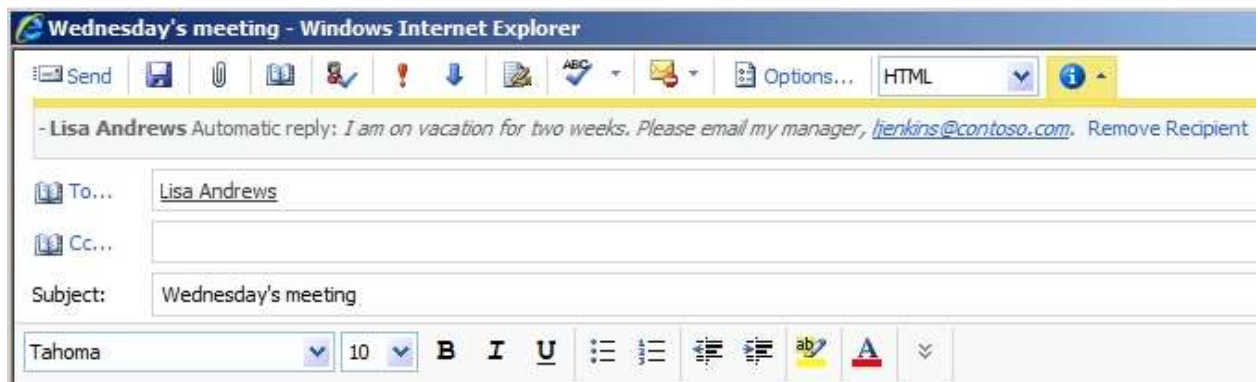

**Figure 2.** Outlook shows an automatic "out of office" note for recipients of your message once you have entered their email address and before composing and sending the message.

Yet, despite the agreement about the importance of the CP and its relative comprehensiveness, Grice has recognized that the goal of "maximally effective exchange of information" is too narrow and that there are maxims that are not covered by the CP such as "Be polite" (Grice, 1975; p 47). Indeed, Grice's work has spawned abundant research on politeness. One of the earliest and most influential works is Lakoff's (1973), which we discuss next.

**3.2 Lakoff's Politeness Rules**

Lakoff's approach includes two major components, or rules, of "pragmatic competence" (Lakoff, 1973). The first rule, "Be clear" relies on Grice's CP. However, Lakoff suggests that effective conversations should go beyond the assumed rationality and the efficiency-based criteria expressed by the CP, as politeness is "a system of interpersonal relations designed to facilitate interaction by minimizing the potential for conflict and confrontation inherent in all human interchange" (Lakoff, 1990; p.34). Thus, a second rule, "Be polite," is required, which consists of a set of three sub-rules: "Don't impose", "Give options", and "Make audience feel good - be friendly" (Lakoff, 1973).

The first politeness sub-rule, *Don't impose*, states that one should not intrude into "other people's business" (p.298). At least, one should ask permission before intruding or dealing with someone else's property. Systems that impose may offend users. Consider working on your computer and then, all of a sudden, it stops working and displays a message indicating that an automatic software update is in progress and a restart is required. Other common examples are unsolicited windows that pop up while browsing internet sites and software packages that by default add unnecessary plugins to your computer during installation, e.g., Adobe's default installation of McAfee (Figure 3). The problem is that many unsuspecting users perform update installations automatically only to later realize that their system was invaded by unwanted software. The pain of dealing with imposing systems offends users so much that some of them choose to install extensions[2] that automatically uncheck unrelated offers when installing or updating systems.

This rule becomes even more critical in the context of people communicating via a CMC system. For example, a social network that automatically notifies other users that you have looked at their profile may be considered impolite because it reveals information about you without asking for your permission.

---
[2] https://unchecky.com/

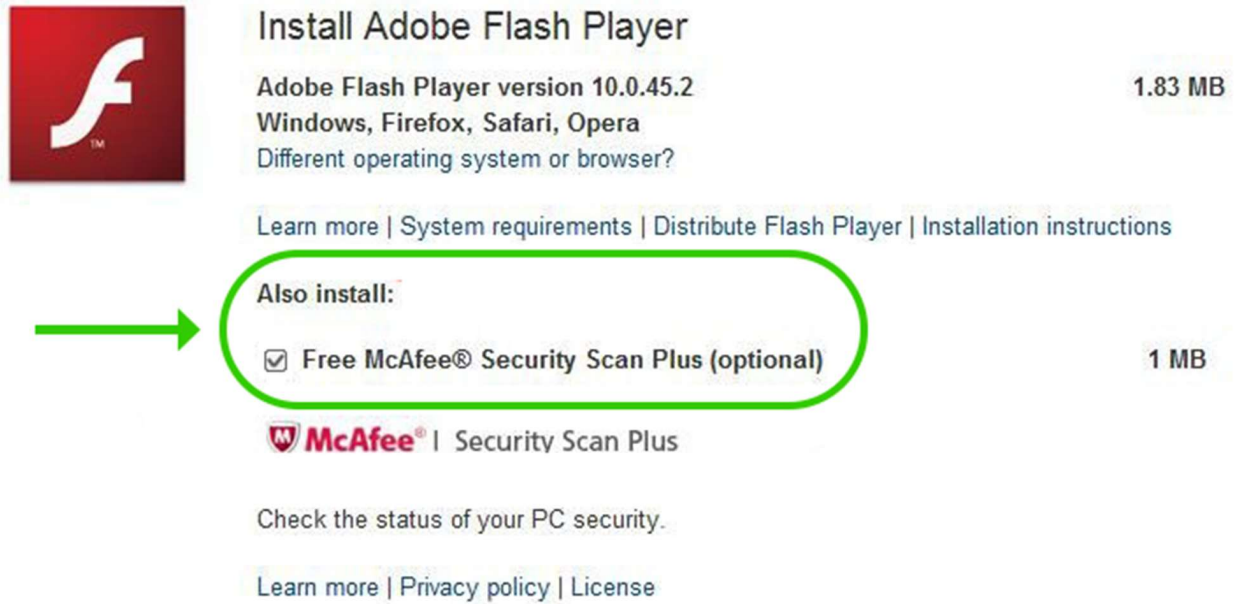

**Figure 3.** Adobe's default installation of McAfee. Users have to uncheck the optional McAfee installation every time they update the Flash Player, otherwise, they end up with an unwanted program that is hard to remove later.

The second rule, *Give options*, states that we give the other person options either to refuse or to accept our requests and desires. Linguistically, it is characterized by saying things hesitantly, leaving the option of a decision to the hearer (Lakoff, 1973). Considering again the automatic software update, it is not only important to ask first for user's permissions, but also to provide options (e.g., Figure 4). A polite software should respect the fact that different users have different preferences. In computer-mediated communication, lack of choice options because of limited communication channels causes frustration. For many years, Facebook provided users with only a "like" symbol in order to react to another user's post. However, this was later redesigned to give users more options with Facebook Reaction, which supports additional degrees of expressing emotions.

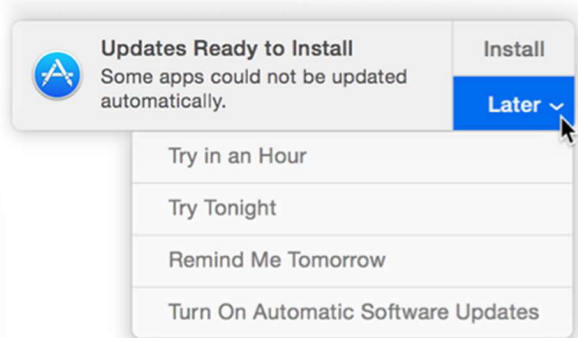

**Figure 4.** Mac notifies when updates are available for the operating system. If it is not a convenient time to install your updates, it gives options to choose a different time.

The third rule, *Be friendly*, emphasizes equality and closeness between the speaker and the hearer. The rule suggests that informal expressions communicate feelings of solidarity that make the addressee feel wanted. Recently, being friendly has become more common in software products. For example, web applications choose a friendly character to represent their brand (e.g., Trello and Mailchimp) and 404 error message (page/file not found) are becoming less formal (e.g., "oh snap!", "Woopsie Daisy", "Our page lost in the sea", see Figure 5). By using these and similar messages ("oops! we're sorry, there seems to be a problem with that page"), the developers make the users feel good by taking responsibility for interaction mishaps). In CMC, supporting friendliness is even more vital. For example, limiting the communication to only formal typing-based communication channels constrain the user's ability to express emotions. Can we imagine a chatting tool without an emoticon keyboard? Today more than 2,500 emoji characters are supported to allow users to express their feelings (Emojipedia, 2017). When people communicate, they use a rich arsenal of words and gestures to behave politely. Interactive technology should be equally rich to support the same purpose.

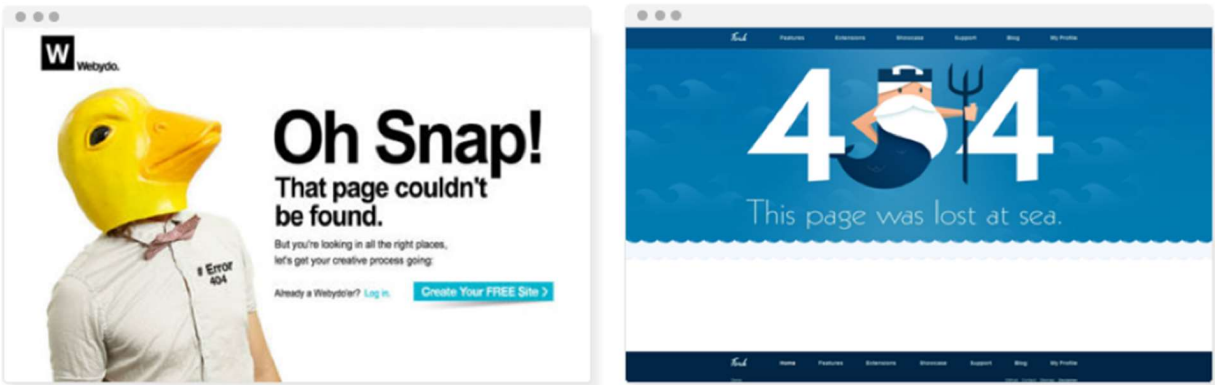

**Figure 5.** Design examples from Webydo.com and Fork-CMS for friendly page-not-found messages.

## 4. PROPOSED RESEARCH FRAMEWORK

Our review of the literature suggests that the concept of politeness in HCI is still at an early stage of development and that it can benefit from additional conceptual and empirical work. Sociolinguistic theories offer rich and multifaceted treatment of the subject. We suggest that Grice's CP and its underlying rules (Grice, 1975) and Lakoff's rules of politeness (1973) can serve as a fertile framework for politeness research in the various domains of HCI. Whereas Reeves and Nass (1996) relied solely on Grice's CP for the conceptualization of politeness, we find value in distinguishing between the CP -- for which, following Lakoff (1973), we use the term "clarity" in our framework -- and Lakoff's further elaboration on politeness, to which we refer as "politeness."

We thus propose a conceptual model (Figure 6) as a first step towards analyzing, designing, and evaluating polite interactive technologies. The model starts with design aspects of a specific system or computerized agent (e.g., interface, interaction and message design), which determine the system or the agent's behavior. Obviously, system behavior consists of various design aspects. The proposed framework focuses on politeness and clarity and therefore it is restricted design aspects that belong to the "Be clear" principle (i.e., the sub-rules that comprise Grice's CP) and those that belong to the "Be polite" principle (i.e., Lakoff's sub-rules). The model suggests that a system designed with such principles embedded in its behavior will influence users' perceptions of the technology's clarity and politeness. However, such influence is not guaranteed. It requires empirical validation and poses various interesting research questions and challenges. Next, these perceptions are translated to higher-order evaluation of the interaction in terms of the focal constructs of our study – politeness and clarity: Does the robot exhibit polite behavior? Does the application communicate clearly its state and its messages to the user? Finally, we suggest that users' sense of politeness and clarity may influence additional perceptions of the interaction, such as enjoyment, trust, and satisfaction. For example, we expect both perceived clarity and perceived politeness to positively affect user satisfaction. Each of perceived clarity and politeness may be associated to different degrees with various UX constructs. Thus, it is likely that as a system follows the rules of clarity to a greater extent, so will users perceive it as easier to use. Or, perceiving the system to be polite is likely to increase users' trust in it. We also expect both perceived clarity and perceived politeness to positively affect user satisfaction. At this stage, we do not aspire to cover the entire nomological network of perceived politeness, and thus we tentatively raise the possibility that it may be related to additional constructs (this idea is marked with "…" in Figure 6).

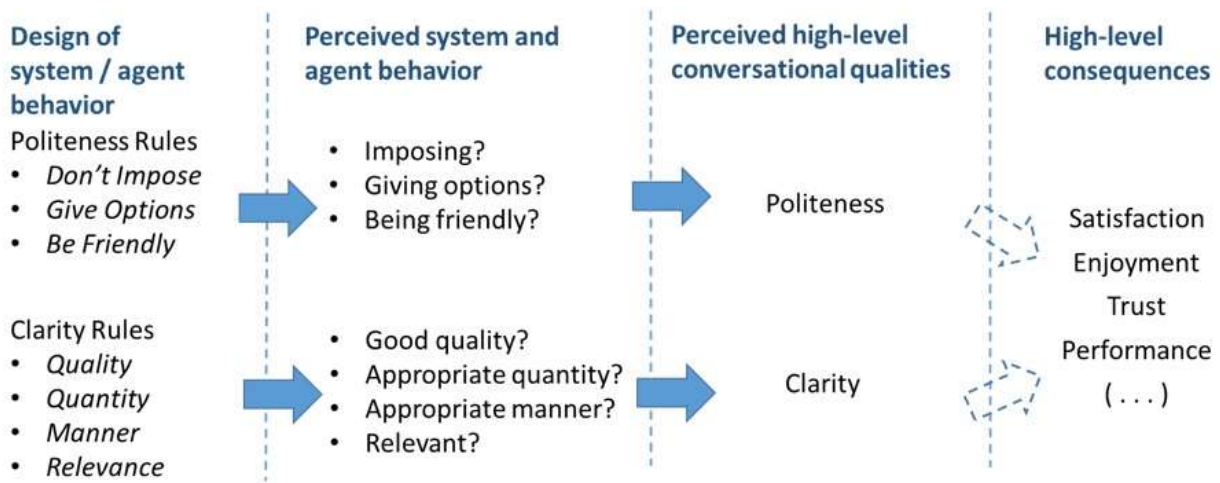

**Figure 6.** A framework for the study of politeness in various HCI domains. The model conceptualizes design properties in terms of Grice's and Lakoff's rules, suggests that they influence users' perceptions of politeness and clarity and that these constructs, in turn, are likely related to other constructs of interest to HCI research.

Previous work in HCI has adopted linguistic theories for research and design (e.g., Winograd and Flores, 1986). However, translating sociolinguistic work poses the challenge of putting philosophical ideas, language-based rules, and anecdotal evidence to empirical tests in the target domain. Indeed, some of the major challenges of this research program lie in the relatively abstract nature of the academic discussion of politeness, its deviation from the intuitive meaning of politeness in everyday parlance, and the lack of previous research to operationalize the concept. The examples of system behavior that follow politeness rules or that deviate from those rules (see Section 3), indicate that the rules that comprise politeness and clarity can be instantiated in various HCI domains. However, we are looking for a more systematic investigation that goes beyond the anecdotal. Moreover, an experimental demonstration that manipulations of politeness rules are perceived by users and, in turn, influence users' evaluations of the system's politeness, will strengthen the evidence for the proposed framework. For this purpose, we have conducted an experiment designed to evaluate the effects of manipulating politeness rules on users' perceptions of system politeness in the context of a computer-mediated collaborative game. The study is described in the next section.

## 5. EXPERIMENTALLY TESTING THE RESEARCH FRAMEWORK

In this study, we examine the research framework (Figure 6) in the context of computer-mediated interaction. For the experiment, we developed a collaborative gaming system, which allowed us to manipulate design aspects of the system and to test whether the manipulations influenced users' perceptions both at the low level (i.e., whether system behaviors induced by the design features have been observed) and at the higher level (i.e., whether the behaviors have been evaluated as clear and polite).

*Hypotheses and Expectations*
This study tests three main hypotheses that stem from the research framework.

*H1*: Following the politeness rules will improve users' perceptions of the system's and of their collaborators' politeness.

*H2*: Following the clarity rules will improve users' perceptions of the system's and of their collaborators' clarity.

*H3*: Following politeness and clarity rules will improve users' experience (broadly defined) from the interactions with the system.

In addition to testing these hypotheses, we have examined other aspects of politeness in computing for which we had no a priori expectations. In particular, we had no empirical or theoretical basis to assume any effects of politeness and clarity on actual performance and on users' perceptions of their performance. Thus, tests of these aspects should be regarded as exploratory.

### 5.1 Method

*5.1.1 The game*
We developed a gaming software based on the Peekaboom game (von Ahn et al., 2006). The software's interface language was Hebrew. In this game, two players take different roles as depicted in Figure 7. The participant in Boom's role starts with an image and a target word that describes some aspect of the image. The participant in Peek's role starts with the same image as Boom's and a list of 20 possible words, only one of which is the target word that appears on Boom's screen. However, the image on Peek's screen is blanketed out and has to be exposed gradually trial by trial. After each incremental exposure, Peek guesses which is the target word until he or she makes the correct guess. Following each guess, Boom provides Peek feedback regarding the correctness of the guess. The pair's mutual goal is for Peek to find the target words with the least number of trials. In this study, each pair of participants played three rounds of the game. The game starts with Peek's turn. Peek reveals square areas of 50 pixels' edge around the click and sends a guess (i.e., a word from the 20-word bank) to Boom. If Peek's guess is correct, both players receive a system message that announces that the guess was correct. They then continue to a new round of the game until they have completed three rounds.

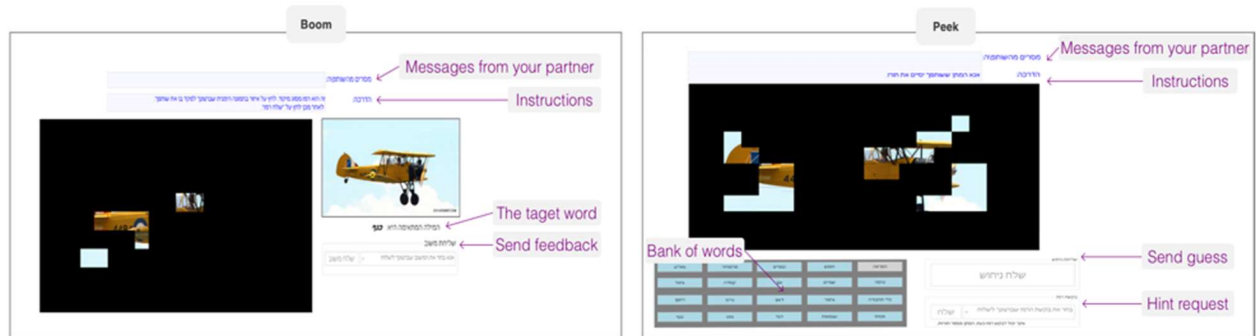

**Figure 7.** Participants' interface for the game. Left: Boom's screen includes the full target image along with the target word and an updated view of Peek's image. Right: Peek needs to uncover a blanketed image and to guess the target word out of a bank of possible words (at the bottom left side of Peek's screen). Peek's guesses and Boom's replies appear as messages at the top of each partner's screen. Reminder: The gaming software's interface is Hebrew.

If Peek's guess is wrong, Boom sends Peek feedback from a list of options. The feedback content is manipulated according to the different conditions of the game as will be explained below. After five trials, Peek's option to request a hint from Boom is enabled. We manipulated the clarity and the politeness of the hint requests according to the experimental condition. These manipulations are described in Section 5.1.2. Upon receiving Peek's hint request, Boom can decide whether to approve or deny the request. In the instructions before the game, we presented the hints as a benevolent act that cannot adversely affect the pair's achievement. To not let strategic considerations enter the participants' interactions, which could have confounded the effects of clarity and politeness, we did not reward or penalize requests for hints or for approval or denials of such requests.

The game included two types of target words. Object-type target words refer to a specific object in the image (e.g., "Tree"). Concept-type target words, on the other hand, describe an abstract concept or a description of a full scene related to the image (e.g., "Party" or "Happiness"). In the first and the third rounds of the experiment, the game included object-type target words. In the second round, the target word was concept-type.

For each type of target word, the system provided a different type of hint (see Figure 7, left). For object-type target words, the hint was graphical. It focused Peek on a small area in the vicinity of the target object (regardless of whether this area was already exposed by Peek or not). For concept-type target words, the hint was provided as a text description. For example, on the right-hand side of Figure 8, the target word is 'Competition' and the textual hint is 'A sporting event with many participants, which will end up with a winner.'

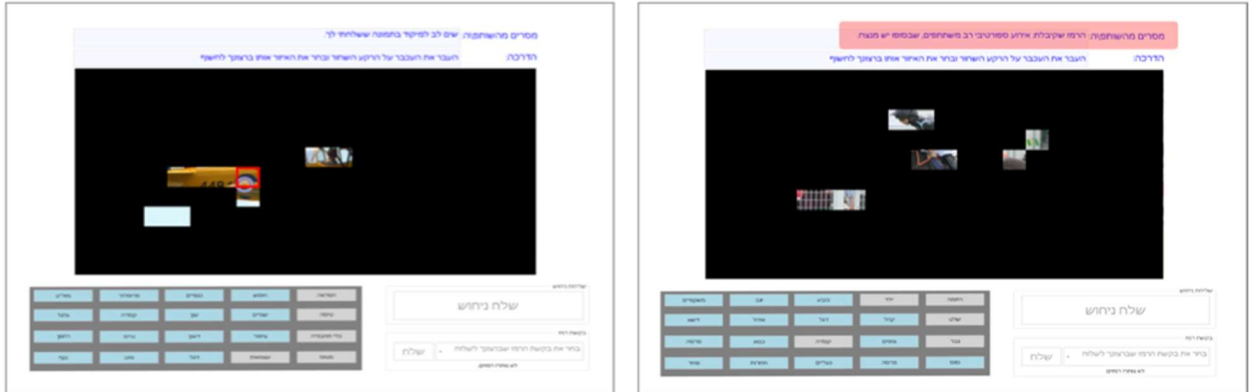

**Figure 8.** Two types of hints as seen from Peek's screen. Left: Boom focuses Peek on a specific area in the target object (red rectangle). Right: a textual hint, seen at the top of Peek's screen (here marked by a pink background).

### 5.1.2 Experimental design and manipulations

The experiment used a 2x2 between-groups factorial design. One factor, politeness, had two levels, low and high. The levels were manipulated by the presence or absence of two of the three politeness rules: Give Options and Be Friendly. The other factor, clarity, also included low and high levels. It was manipulated based on three of Grice's four conversational rules; Quantity, Manner, and Relevance. We did not manipulate the fourth rule, Quality, since we estimated that this would have too strong an effect on Peek's ability to guess the target word and the pair's evaluations of the system and of their partner. The manipulations were based on correspondences between the players, an idea inspired by an educational software for language tutoring presented in (Johnson and Wang, 2010). The manipulations are elaborated below and summarized in Tables 1 and 2.

|  | **The content of the feedback from Boom to Peek** | **The content of the hint request from Peek to Boom** | **The content of Boom's rejection of Peek's hint request** |
|---|---|---|---|
| Politeness rule used | Be Friendly | Be Friendly + Give Options | Be Friendly + Give Options |
| Condition: High | Boom can select the feedback content from a list of three options. Peek receives the selected message and a friendly emoji that is added automatically to the message. | Peek can select the content of the hint requests from a list of four options. | Automatic hint rejection with a friendly tone ("Sorry, a bit too soon for a hint, try again later"). |
| Condition: Low | Boom cannot select the content of the feedback. There is only one option: "You're wrong, your guess is incorrect". | Peek cannot select the content of the hint requests; There is only one option: "Send me a hint". | Automatic, not friendly, hint rejection ("I do not approve your request. Try later"). |

**Table 1.** Manipulations of the two politeness experimental conditions (High vs. Low).

|  | **How Boom sees Peek's guess - Textual** | **Hints - Textual** | **Hints - Graphical** |
|---|---|---|---|
| Clarity rule used | Quantity | Relevance | Manner |
| Condition: High | Your partner guess is <*guess*>. This guess is incorrect; please send him feedback. | Predefined textual hint based on the image and the target word. | A circular outline of radius = 20 pixels around the clicking point. |
| Condition: Low | Your partner guess is <*guess*>. This is guess number <*number*> and it was send in <*timestamp*>. This guess is incorrect because the correct word is <*word*>. Now you need to send feedback about this guess to inform your partner that this guess is incorrect. | Constant hint that repeats the instructions but is not specific to the target word. | A circular outline of radius = 60 pixels around the clicking point |

Table 2. Manipulations of the two clarity experimental conditions (High vs. Low).

*Manipulating Politeness*
To manipulate politeness, we operationalized the Give Options and the Be Friendly rules (see Table 1). In the Polite-High condition, messages between the two players were structured such that they had several optional phrases to select from. In addition, a friendly emoji was added to Boom's messages to Peek to reflect a friendly approach. Similarly, if Boom rejected Peek's request for a hint regarding the target word, the rejection message was phrased in a friendly manner. In the Polite-Low condition, the players could only use one predefined message, and the hint rejection was phrased in a non-friendly manner.

*Manipulating Clarity*
The issue of message clarity has been identified from the early days of the HCI field as a crucial element for users' understanding of interactive systems (e.g., Shneiderman, 1982; Lewis and Norman, 1986). In this experiment, we manipulated three aspects of clarity either textually or graphically (Table 2). We used textual hints to display Peek's guess to Boom. In the Clarity-High condition, the guess was seen as a concise message indicating Peek's guess, whether it is correct, and a reminder to Boom to send feedback. In the Clarity-Low condition, the message included unnecessary information that did not help in clarifying the communication or possible acts by Boom.

For hints related to the concept-type words, Boom gave predefined, relevant textual hints that describe the target word. For example, if the image in the game was of kids playing, then possible words in the word bank could have been 'childhood,' 'fun,' 'happiness,' and 'competition'. Suppose the target word was "fun,' then the hint would be "behavior or activity that is intended purely for amusement".

For hints on object-type words, we used a graphical manipulation, in which Boom could point Peek towards the target word by "pinging" a specific part of the image. This was done by clicking on a part of the image. In doing so, Boom helps to clarify which are the most relevant parts of the image for the target word. Clarity was manipulated here by the rule of Manner. This was operationalized as the number of pixels of the "pings." For example, if two players were playing with an image of an elephant, then possible words in the word bank could have been 'elephant', 'trunk', 'tusk', 'ear'. Suppose the target word is 'trunk', then Boom can "ping" an area on the trunk of the elephant by clicking on it, and focusing Peek on this part. For the Clarity-High condition, the ping area was focused with a radius of 20 pixels to avoid ambiguity. For the Clarity-Low condition, the ping was less focused, with a radius of 60 pixels, allowing for a more ambiguous interpretation of the hint.

*5.1.3 Procedure*

Participants were seated simultaneously in two separate computer laboratories in groups of up to 10 participants per room per session. In each room, all participants assumed the same role (of either Peek or Boom). Following a short introduction by the experimenter, the participants were asked to read detailed game instructions about the game's flow and their specific role in the game. After reading the instructions and signing a consent form, they started the game, completing all three rounds of the game in succession. After finishing the third round, the participants completed a questionnaire of 14 items, one of which was a quality control item ("This is a control question, please choose 5 as an answer to this question"). The other items queried the participants about their experience with the game's application and their perception of the other player. On average, the experiment lasted 17 minutes.

*5.1.4 Images*

The experiment used a pool of 40 images (24 images for the object-type rounds and 16 images for the concept-type round). For each image, a list of 20 words was created, including the target word. The images displayed to each pair of participants were randomly selected without replacement from the image pool.

*5.1.5 Participants / sample*

One hundred and twelve third-year engineering students participated in the experiment for class credit. The participants were not exposed during their academic studies before the experiment to politeness or clarity considerations of software applications. A pair of students who could not complete the experiment due to technical issues were excluded from further analyses. Of the remaining 110 participants, 39 were females and the age range was 20 to 29 (average = 24.8).

*5.1.6 Dependent Variables*

The main variables of interest in this study are subjective evaluations by the participants of their achievement in the game, and of the gaming system's and the other player's behavior. We also collected a set of objective performance measures of the pairs of players.

The objective variables were the length of the game (i.e., the total number of guesses until Peek has found the target word), total hints supplied (i.e., the total number of hints that Boom supplied), and the total number of hints rejected (i.e., the total number of the hints requests that were rejected by Boom).

The subjective evaluations were based on the post-experimental questionnaire. The questionnaire included 14 items: four manipulation check items and ten items regarding various aspects of their experience playing the game. The aspects measured by the items are perceived politeness, perceived clarity, ease of use, enjoyment, satisfaction with the gaming software, and perceived performance. All of the items included a 7-point response scale ranging from "strongly disagree" to "strongly agree". We did not use multiple item scales because we were concerned that they would wear down the respondents. While using single-item scales is not recommended in general, the weakness of this approach is mitigated by the fact that the constructs measured are relatively uncomplicated and are at least apparently unidimensional. In such cases, the use of single-item scales is acceptable (cf. Gardner et al. 1998; Bergkvist, 2015; Ang and Eisend, 2018). Thus, we measured the various aspects by either using a single item or two items. Each of the politeness and clarity aspects was measured by two different items, one relating to the behavior of the gaming system and one relating to the behavior of their partner. For this reason, they could not be combined into a single scale. Satisfaction and perceived performance were measured by two similar items, which were proven to be highly correlated (r = 0.75 and r = 0.79, respectively), and were therefore averaged for a single score.

For manipulation checks, we measured the participants' perceptions of the manipulations. Thus, items regarding the manipulations of politeness asked the participants whether the messages they received were friendly and whether the system allowed them optional messages to choose from. One item checking the manipulation of clarity was presented to Boom regarding the level of detail of information included in Peek's guess. Another item asked Peek about the relevance of Boom's hints. All of the questionnaire's items are presented in Table 3, grouped by topic (in the experiment items were presented in mixed order). Items' means and standard deviations and correlations between items are provided in the Appendix.

| Aspect Measured | Item code and text |
| --- | --- |
| Manipulation check - politeness | MCP1. The system allowed me to select among messages to my partner |
| | MCP2. The messages I received from my partner were friendly |
| Manipulation check - clarity | MCC1. The messages I received about my partner's guesses were at the right level of detail |
| | MCC2. The hints I received from my partner were relevant |
| Politeness | P1. The software allowed me to be polite towards my partner |
| | P2. The messages I received from my partner were polite |
| Clarity | C1. The messages I received from my partner were clear |
| | C2. The gaming software enabled me to send exactly the information I wanted to communicate |
| Enjoyment | E1. I enjoyed the cooperation with my partner |
| Ease of use | EOU1. It was easy to use the gaming software |
| Satisfaction * | S1. I am satisfied with the way that the gaming software allowed me to communicate with my partner |
| | S2. In general, the gaming software supports good communication between people |
| Perceived performance * | PP1. On average, our team discovered the picture descriptions very fast |
| | PP2. Relative to other pairs, our performance was very good |
| Quality control | QC1. This is a quality control item; please choose 5 as an answer to this item |

* Both items measuring this aspect were averaged to form a single score.

**Table 3:** Post-experiment questionnaire items

### 5.2 Results

The experiment's main objective was to study the participants' reactions to the manipulations of the software. Therefore, we focus first on the analysis of the post-experiment questionnaire, which measured the participant's subjective evaluations regarding various aspects of the game. Of the 110 participants, five did not answer correctly the quality control item in the questionnaire and were excluded from the analysis. Thus, the analysis of the questionnaire's item relies on data obtained from the other 105 participants. We begin with checks of the experimental manipulations and then continue to the participants' perceptions of higher-level constructs such as politeness, clarity, satisfaction, and enjoyment. All of the questionnaire's items are listed under their relevant subsections below.

*5.2.1 Manipulation Checks*
We tested whether the manipulations described in Tables 1 and 2 were perceived by the users. The measures were analyzed as dependent variables in 3-way between-group analyses of variance (two levels of clarity, two levels of politeness, and two levels of role - Peek or Boom). The results indicate that the manipulations of the politeness rules – Be Friendly and Give Options – were successful. The participants in the Polite-High condition agreed more ($M = 3.44$, $SD = 2.24$) with the item that checked the Give Options manipulation (MCP1) relative to the participants in the Polite-Low condition ($M = 2.24$, $S.D. = 2.0$). The difference was statistically significant ($F(1, 97) = 8.46$, $p = .005$, *partial* $\eta^2=.08$)[3]. The Polite-High participants were also more likely to agree with the manipulation check of the Be Friendly manipulation (MCP2) ($M = 6.43$, $S.D. = 0.98$) relative to the Polite-Low condition ($M = 5.63$, $S.D. = 1.69$). This difference was similarly significant ($F(1, 97) = 8.13$, $p = .005$, *partial* $\eta^2=.08$). There were no other main, or interaction effects on these items.

However, the manipulations of the clarity rules were not successful. There were no significant differences between the Clarity conditions in terms of the items that checked the manipulation of the Quantity rule (MCC1) and the manipulation of the Relevance rule (MCC2). In both cases, however, the average response was above the scale's average $M = 4.70$, $SD = 2.21$ for MCC1, and $M = 5.69$, $SD = 1.67$ for MCC2), indicating that clarity was not a major issue during the game. In addition, there were no other main, or interaction effects on these items.

*5.2.2 Perceived Qualities of the gaming system*
Commensurate with the results of the manipulation checks, no significant main effect of clarity was found on any of the experience items. Thus, the rest of the analyses use 2-way analyses of variance with politeness (high, low) and role (Peek, Boom) as the experimental factors. We present the results of this section by the aspects measured in the questionnaire.

*Politeness*
Overall, the participants rated the system as polite. On a 1-7 scale, the average response to item P1 was 5.96 ($SD = 1.53$). The participants also tended to agree with P2 ($M = 6.10$, $SD = 1.39$). A 2x2 analysis of variance with politeness (high, low) and role (peek, boom) as between-group factors and P1 and P2 as dependent variable revealed a significant effect of politeness ($F(1, 101) = 12.07$, $p = .001$, *partial* $\eta^2=.11$ and $F(1, 101) = 14.10$, $p < .001$, *partial* $\eta^2=.12$, respectively). The main effect of politeness on P1 was qualified by a Politeness x Role interaction ($F(1, 101) = 7.79$, $p = .006$, *partial* $\eta^2=.07$). This interaction, depicted in Figure 9, indicates that participants in Boom's role experienced the system in the Polite-Low condition as providing less support for being polite towards their partners (*M* = 4.77, *SD* = 2.09) relative to the Polite-High condition (*M* = 6.46, *SD* = 1.17).

---

[3] It can be argued that analyzing single-item scales using parametric tests is problematic due to the ordinal nature of such scales. However, Norman (2010) suggests that parametric methods are robust for single-item scales of the kind used in this study. Non-parametric tests, on the other hand may have their own shortcomings, such as inability to test interactions or to provide unbiased test results (Lupsen, 2017, 2018). Nevertheless, we ran non-parametric analyses (Mann-Whitney-Wilcoxon) on all main effects, which yielded very similar results to the parametric tests reported here.

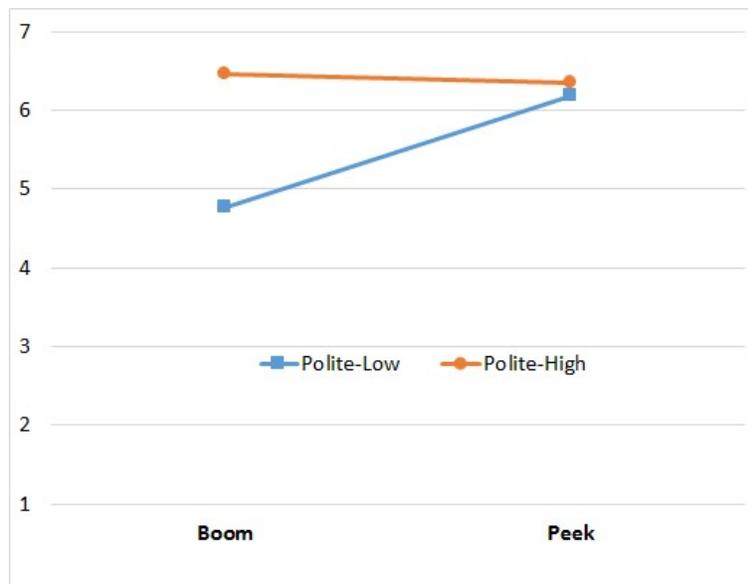

**Figure 9.** Average participants' response to the item "The software allowed me to be polite towards my partner."

*Clarity*

The two items measuring the participants' perceptions of clarity (not to be confused with the manipulated independent factor of clarity), were analyzed by a two-way ANOVA. There was no significant effect on the C1 item. However, there was a Role effect on C2 ($F(1, 97) = 21.66.31$, $p < .001$, *partial η²*=.18), which was qualified by Politeness x Role interaction effect ($F(1, 97) = 4.31$, $p = .04$, *partial η²*=.04). Participants in Peek's role perceived the system as allowing clearer communication, especially in the Polite-Low condition (*M* = 3.88, *SD* = 1.86, compared to *M* = 2.85, *SD* = 2.07 for Boom) and even more so in the Polite-High condition (M = 5.27, *SD* = 1.91, compared to *M* = 2.68, *SD* = 2.11 for Boom), as depicted in Figure 10.

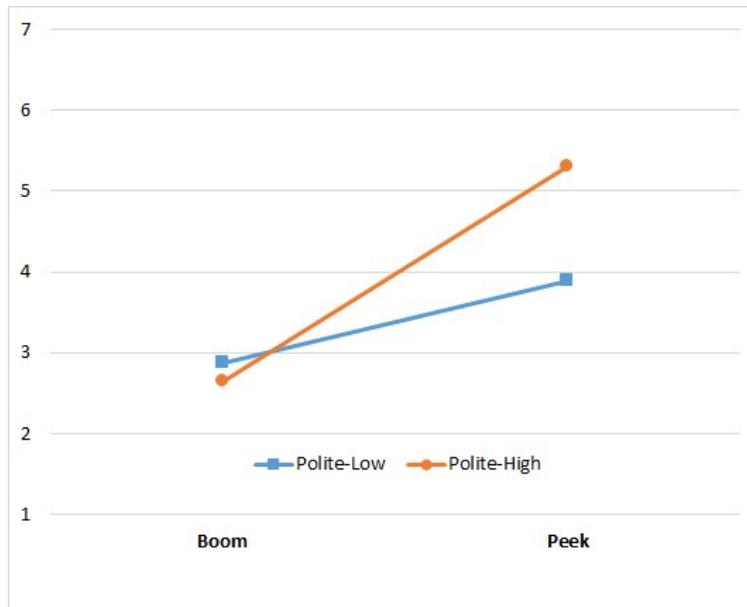

**Figure 10.** Politeness x Role interaction for the item "The gaming software enabled me to send exactly the information I wanted to communicate."

*Overall Evaluations*

In addition to perceived politeness and clarity, we measured other, potentially related aspects of the participants' experience using the gaming system.

The system was easy to use (Item EU1, Mean = 6.24, *SD* = 1.21), which may have created a ceiling effect that led to a lack of statistically significant differences between groups. However, participants' enjoyment from cooperating with their partners (Item E1) was influenced by the system's politeness level. A two-way ANOVA with politeness (high, low) and role (Peek, Boom) as between groups factors found a significant effect of politeness on enjoyment ($F(1,101) = 8.27$, $p = .005$, *partial* $\eta^2=.08$). Participants in the Polite-High condition enjoyed the cooperation more than participants in the Polite-Low condition (*M* = 5.28, *SD* = 1.76 vs. *M* = 4.24, *SD* = 1.96).

The two items that measured the participants' satisfaction were highly correlated ($r = .75$) and were combined to form a single satisfaction score. The 2X2 ANOVA revealed a main effect of Role on satisfaction ($F(1,101) = 13.92$, $p<.001$, *partial* $\eta^2=.12$). Participants in Peek's role were more satisfied (*M* = 4.44, *SD* = 1.63) than participants in Boom's role (*M* = 3.16, *SD* = 1.85).

*Perceived vs. Actual Performance*

The median number of trials (guesses) needed by the pairs to complete the three rounds of the game was 23.0 (*M* = 23.1, *SD* = 7.49 *min* = 9, *max* = 44). The median number of trials needed in Rounds 1 and 3 (in which the target word was an object-type) was 8.5 and was 5 in Round 2 (in which the target word was a concept-type). The median number of hints supplied during the three rounds was 2 per pair (*M* = 1.98, *SD* = 0.91). Almost no hint requests (from Peek) were rejected (by Boom) during the game (*M* = 0.17, median

= 0, max = 2). A two-way analysis of variance with politeness (high, low) and clarity (high, low) found no statistically significant effects of those experimentally manipulated factors on the performance measures.

The two items that measured the participants' perceptions of their performance (PP1 and PP2) were highly correlated ($r = .79$) and were combined to form a single perceived performance score. Participants in the Polite-High group perceived their performance as better than participants in the Polite-Low group ($M = 4.03$, $SD = 1.39$ vs. $M = 3.51$, $SD = 1.47$, respectively). However, there was a moderate correlation between perceived performance and actual performance, as measured by the number of trials it took to finish the game ($r = .42$, $p < .01$). Thus, to test whether perceived performance was influenced by any of the experimental factors we conducted a 3-way ANCOVA with politeness, clarity, and role as between groups factors, and the number of trials needed to finish the three rounds as a covariate. Only the covariate (performance) was significant in the analysis ($F(1, 96) = 18.57$, $p < .001$, *partial η2*=.14), indicating that perceived performance was moderately calibrated with actual performance but not with any of the independent factors. There were no other statistically significant main- or interaction effects.

**5.3 Discussion of experimental findings**

As a first step towards empirical examination of the framework, we designed an experiment in which politeness-related design features can be manipulated and controlled. The experiment was conducted in the context of computer-mediated collaborative gaming.

The results indicate that manipulations of two politeness rules – Give Options and Be Friendly – had been perceived by users and had the expected effect on users' perceptions of their partners' and of the software's politeness. This finding provides empirical support to relatively abstract linguistic politeness rules, to the appropriateness of those rules to the area of CMC, and to the prospects of grounding research on politeness in HCI on theoretical foundations of sociolinguistics research.

Moreover, in addition to demonstrating the effects of design features on perceptions of politeness, the experiment has shown that the more polite system also had a positive effect on users' enjoyment from their cooperation with their gaming partner and on their feeling of being able to communicate better with their partner. One obvious explanation for the latter findings is that politeness has a role in streamlining social exchange, which increases enjoyment. This should not preclude other possible explanations such as a halo effect of polite behavior. Such explanations were outside the scope of the study but should be explored in future research.

We also found that the role that participants played in the game interacted with the politeness condition in affecting their perceptions of being able to be polite towards their partners. The source of the interaction was the stronger politeness effect for people in Boom's role. We did not expect this difference in advance, but in retrospect, it seems that there was too little room in general for Peek to request hints (on average, about two hints were requested overall during the three rounds). Therefore, situations in which Peek could choose the message sent to Boom (and behave more politely) were less common. On the other hand, participants in Boom's role had the option to choose the content of the feedback whenever Peek sent them a guess (on average, about 23 times during the experiment). Consequently, the politeness manipulation on Boom's side was more effective. This finding may suggest that multiple opportunities for polite behavior may have a stronger effect on perceived politeness than contexts in which such opportunities are rare.

The politeness effect was statistically and practically significant despite a relatively mild manipulation that did not alter the participants' performance in a one-time, short interactive session. This effect would likely be even stronger under continuous use or under a more consequential context. In this respect, the experiment's results reinforce the claim that politeness is an important ingredient in human-computer interaction.

In contrast to the effects of the politeness manipulation on enjoyment, it did not have a significant direct effect on users' satisfaction. We can offer two possible explanations for this. First, while enjoyment and satisfaction are related, satisfaction can be influenced by additional factors, such as performance, which were not very salient in this study. Thus, the relatively mild manipulation of politeness, which was sufficient to influence enjoyment, was not strong enough to influence satisfaction. Clearly, future studies should further examine the effects of stronger, and perhaps different, politeness manipulations on satisfaction as well as on other relevant constructs. A second explanation is that the effects of politeness on satisfaction are not direct (as suggested in Figure 6) but rather mediated via other constructs, such as enjoyment. Given the nascent and exploratory nature of this research, additional studies are required to explain this result.

### 5.4 Study Limitations

A salient limitation of this study is the failure to manipulate the rules of clarity. A possible cause for the weakness of the clarity manipulation was the between-subjects design used in this study. Using a system for a short time without the ability to compare to similar systems may diminish manipulations of the system's features. However, this limitation has also some unintended benefits. First, it demonstrates the separability of the concepts of clarity and politeness and their associated rules, given that neither manipulation has caused effects in perceptions of the other. This is an important empirical support for our research model. Moreover, it emphasizes the success of the politeness manipulation under relatively unfavorable experimental conditions. This strengthens the claim that politeness rules can be designed into an interactive system and that they affect the perception of politeness and, in turn, of other related constructs, such as enjoyment, as demonstrated in this study, and trust (e.g., de Visser and Parasuraman, 2010).

Another limitation of this study is that it is confined to the context of collaborative gaming. As mentioned above, we suggest that the proposed framework is relevant to a multitude of domains, including ordinary applications, such as user-to-application or user-to-website interaction, social networks, and human-robot interaction. The last two types of domains are particularly intriguing since they represent interactions that may include a major social component. However, the claim for our framework's generalizability should be tested in those other domains as well.

Similarly, the participants in this experiment were engineering students in their 20's. Since perceptions of politeness and clarity in the computing domain may vary due to age, culture, and education (e.g., (Lakoff and Ide, 2005), additional testing of the framework is required that takes into account those moderating factors. Likewise, it is important to mention that Lakoff's politeness theory originated from observations of differences in how women and men communicate. Our design and sample did not allow us to test gender differences in reactions to politeness manipulations, but this issue should be investigated in future studies on politeness in HCI.

# 7. GENERAL DISCUSSION

Aspects of politeness in interactive technology are becoming increasingly relevant with the proliferation of social applications and the introduction of new forms of computing such as robots into the social sphere. Yet, research on politeness in the various domains of human-computer interaction has been sporadic. Empirical studies have used a relatively coarse view of politeness and have tended to focus on people's behavior towards technology (Nass, 2004; Nass and Moon. 2000; Nass et al., 1999). A more refined framework for the study of politeness (Whitworth, 2005) had not been tested empirically. In this paper, we have presented a view of politeness in HCI that borrows from work in the areas of sociolinguistics and pragmatics. The framework provides a more systematic approach to research on politeness in HCI in several ways. First, it delineates the place of politeness in HCI as an important aspect of (a) communication between humans and interactive technology or (b) computer-mediated communication among humans, by comparing it to the important -- yet separate --concept of clarity. Second, the framework relies on explicit rules of communication that we suggest (a) give rise to perceptions of politeness in HCI, and (b) can serve as design principles in interactive technology development. Third, it proposes how the concept of politeness and clarity may be related to other relevant HCI constructs such as satisfaction and ease of use. Moreover, we have empirically tested the framework and have provided evidence for its soundness. Thus, we argue that basic rules that were suggested as governing conversations among people are equally relevant to interactions with computers or computer-mediated interactions.

We consider the empirical evidence provided in this paper promising yet preliminary. We acknowledge the need to replicate our experimental findings in other domains and contexts. Within such efforts, the framework and the empirical findings offer a multitude of implications and potential directions for future research on politeness in HCI and for designing polite interactive technology.

From a practical perspective, the framework suggests three general rules, which software applications, social media, social robots, and other interactive technology should follow. Indeed, the politeness rules are general and thus they provide designers only broad guidelines. Yet, these guidelines are more informative than the vague notion of politeness in computing had been thus far. They also indicate that politeness in HCI should transcend speech to also include behavior. Examples of such behavior that is not imposing or which keeps the users' options open for them are provided in Figures 4 and 5. This aspect of politeness could also be quite salient in asocial robotics where gaze (Stanton and Stevens, 2017), gestures (Pan et al., 2018), and movement (Herrera, 2017) may bear on perceived politeness.

When designing interactive systems, it is tempting to view clarity and politeness as two competing and irreconcilable aspects of communication. Whereas clarity rules emphasize efficiency, politeness rules relate more to interpersonal and social relations. Such a distinction echoes the distinctions between pragmatic and hedonic aspects of system design (Hassenzahl, 2003). However, in reality, these distinctions are not clear-cut. Depending on the circumstances, the two rules may coincide and reinforce each other, or be in apparent conflict in which one or the other supersede (Lakoff, 1973). This means that designers have leeway in determining the behavior of a system along the two aspects, as opposed to necessarily choosing one over the other.

The proposed framework offers multiple research directions. Obviously, social robotics is a natural domain for further research on interactive politeness. Early research in this field suggests that people are sensitive to robots' polite behavior (Inbar and Meyer, 2019; Kato et al., 2015). We should explore ways to endow robots with politeness capabilities that will contribute to the interactive experience, and investigate whether factors such as users' age, robot type and context of operation affect those capabilities.

The empirical support for the role of politeness in HCI begs the question of whether adjacent concepts may also be relevant to the field. For example, in everyday parlance, civility and politeness may be used interchangeably. Yet, these concepts are not the same. Lakoff suggests that "if politeness (whether positive or negative) is an offering of good intentions, civility is a withholding of bad ones, a decision not to do something negative that one might have otherwise done" (Lakoff, 2005, p. 25). Currently, our framework is not sensitive to such distinctions, but future extension and refinement of it may suggest whether considerations of civility should also be taken into account and whether there are advantages in designing for civility that can improve human communication with or via computerized systems.

An interesting question raised by our framework is whether perceptions of politeness and clarity derive directly from the degree to which systems adhere to politeness and clarity rules (the left-hand side items in Figure 6). Other aspects that are not specified as rules in our framework may also affect perceptions of politeness and clarity. For example, Whitworth (2005) suggested openness and being personal as polite computing rules that have no direct counterparts in our framework. Future research may investigate these rules and their effects on perceptions of politeness as well. Furthermore, the proposed framework does not specify which of the politeness rules looms larger in people's overall evaluations of a system's politeness. The answer to this question is likely to depend on domain and context (e.g., the type of technology involved, use setting, the users and their goals) and even on personal preferences and sensitivities. Future studies on politeness should take these potential moderators into account.

More research is needed regarding the pattern of relations between politeness and related constructs, such as those presented on the right-hand side of Figure 6. First, it is important to uncover what constructs other than the ones mentioned above are affected by politeness. Second, it is also likely that the relations between the constructs in this nomological net are moderated by contextual factors such as domain (e.g., application vs. HRI vs. CMC) or culture (e.g., national or age-related). Third, the nature of the relationship between perceptions of politeness and clarity can also be the subject of future research. Whereas both clarity and politeness are principles that facilitate communication, they are different concepts with unclear boundaries. Thus, there is ample room for further exploration of the relations between the two constructs and their relative importance to human-computer interaction. Given that politeness and clarity may represent two competing criteria in certain contexts (for example, when the need to give the user multiple options results in the presentation of too much information), which criterion should prevail? Lakoff and Ide (2005) suggest that "in daily intercourse, when faced with a choice between clarity and politeness, people normally opt in favor of the latter" (p.8). It is our impression that in the various domains of human-computer interaction, the answer to this choice is likely to be more nuanced and context-dependent. Empirical research may test this conjecture in various domains of interactive technology.

Finally, we have chosen the particular framework described in this paper for two major reasons: its simplicity and its relevance to non-verbal behavior. On balance, this approach does not currently support specific politeness guidelines for specific contexts. Brown and Levinson's (1987) highly cited politeness theory may provide more nuanced rules. Yet, those rules are also more complex and more difficult to adopt to non-linguistic behavior. In addition, we suspect that the concept of "face," which is central to Brown and Levinson's theory may be too strong in the context of interacting with technology. Moreover, all politeness theories, including those relying on the concept of face have been criticized on various grounds (Kasper, 1990; Mao, 1994, Eelen, 2014). Hence, a quest for a general politeness theory may be futile. Instead, future research on polite computing may look into the adequacy of various frameworks based on their relative strengths and shortcomings. In our opinion, when adopting or developing such a framework, evaluation of its relative merit should include two important criteria: (1) the framework's potential to explain and predict user behavior, and (2) the simplicity with which the framework can be translated to practical design guidelines for developers.

## 8. CONCLUSION

We propose a conceptual framework, which can be used for studying politeness issues in HCI and for providing guidance to designers of such technologies. The framework seeks to introduce a more disciplined treatment of the topic. It demonstrates how principles of politeness in human conversation can be applied to design principles of interactive systems. An empirical study reported in this paper provides support to the basic tenets of the framework. The results are commensurate with social science findings that politeness has an important role in regulating communication and enhancing social interactions. Taken together, the framework and the empirical studies substantiate the call for the elevation of studying and designing polite computing from a relatively intuitive affair to one based on more systematic principles.

The concept of politeness in the proposed framework is based on three basic rules, which apply not only to the language used by information technology or facilitated by its features but also to its behavior and the behavior of other users which it enables, promotes, or restricts. The framework delineates the important interplay between the concepts of politeness and clarity in human communication. It provides support for the applicability of these concepts to human-computer interactions and means to operationalize, measure, and design politeness in this context. Its principles are general enough to be applied to various domains of HCI, including computer-mediated communication, social applications, and social robotics. Finally, the framework provides a sound starting point to explore relations between politeness and other important aspects of the user experience.

# APPENDIX

**Items' means and standard deviations and correlations between items**

| | Mean (SD) | P1 | P2 | C1 | C2 | PE | EOU | S1 | S2 | Satisfac-tion | PP1 | PP2 |
|---|---|---|---|---|---|---|---|---|---|---|---|---|
| P1 | 5.96 (1.53) | -- | | | | | | | | | | |
| P2 | 6.10 (1.39) | .60** | -- | | | | | | | | | |
| C1 | 4.59 (2.10) | .37** | .33** | -- | | | | | | | | |
| C2 | 3.65 (2.22) | .15 | .17 | .37** | -- | | | | | | | |
| PE | 4.77 (1.92) | .18 | .22* | .47** | .33** | -- | | | | | | |
| EOU | 6.24 (1.21) | .43** | .34** | .26** | .19 | .28** | -- | | | | | |
| S1 | 4.07 (2.16) | .22* | .16 | .47** | .67** | .36** | .25* | -- | | | | |
| S2 | 3.50 (1.80) | .25* | .21* | .50** | .68** | .41** | .17 | .75** | -- | | | |
| Satisfaction ^ | 3.78 (1.85) | .25* | .19* | .51** | .72** | .41** | .22* | .95** | .93** | -- | | |
| PP1 | 3.80 (1.55) | -.03 | -.02 | .41** | .24* | .41** | .06 | .16 | .25** | .22* | -- | |
| PP2 | 3.75 (1.51) | -.05 | -.08 | .33** | .23* | .39** | -.02 | .15 | .24* | .21* | .79** | -- |
| Perceived performance^^ | 3.78 (1.45) | -.05 | -.05 | .39** | .25* | .42** | .03 | .17 | .26** | .22* | .95** | .94** |

* p<.05;  ** p<.01;  ^ Average of S1 and S2;  ^^ Average of PP1 and PP2

N = 105